\documentclass[10pt,twocolumn]{article}
\usepackage[top = 1in,bottom = 1in,left = 1in,right = 1in]{geometry}
\usepackage[cmex10]{amsmath}
\usepackage[pdftex]{graphicx}
\usepackage{multicol}

\title{Airport Gate Scheduling for Passengers, Aircraft, and Operation}
\author{Sang Hyun Kim, Eric Feron\footnote{Eric Feron is also with \'Ecole Nationale de l'Aviation Civile.}, John-Paul Clarke \\ School of Aerospace Engineering \\ Georgia Institute of Technology \\ Atlanta, GA, U.S.A. \\ sanghyun.kim, feron, johnpaul@gatech.edu
\and
Aude Marzuoli\footnote{Aude Marzuoli is also with Georgia Institute of Technology.}, Daniel Delahaye \\ \'Ecole Nationale de l'Aviation Civile \\ Toulouse, France \\ aude.marzuoli, delahaye@recherche.enac.fr
}

\begin{document}

\begin{multicols}{2}
\maketitle
\end{multicols}

\begin{abstract}
Passengers' experience is becoming a key metric to evaluate the air transportation system's performance. Efficient and robust tools to handle airport operations are needed along with a better understanding of passengers' interests and concerns. Among various airport operations, this paper studies airport gate scheduling for improved passengers' experience. Three objectives accounting for passengers, aircraft, and operation are presented. Trade-offs between these objectives are analyzed, and a balancing objective function is proposed. The results show that the balanced objective can improve the efficiency of traffic flow in passenger terminals and on ramps, as well as the robustness of gate operations. 
\end{abstract}

\section{Introduction}
Flight delays do not accurately reflect the delays imposed upon passengers' full itineraries. The growing interest to measure the Air Transportation System's performance calls for new metrics, reflecting passengers' experience \cite{cookpassenger}. The cost of congestion and delays in such a complex system is huge. In 2011, according to Airlines for America, 103 million system delay minutes have cost \$7.7 billions to scheduled U.S. passenger airlines \cite{airlines}. Because of the hub-and-spoke structure of the airports network, major airports, such as Hartsfield-Jackson Atlanta International Airport, have a significant impact on the performance of the overall system. In particular, connecting passengers in such hubs may represent the largest share of traffic and are most vulnerable to delays that can severely perturb their journeys. In worst cases scenarios, a single delay can "snowball" through the entire network \cite{ahmadbeygi2008analysis}.

Airport Collaborative Decision Making (A-CDM) aims at reducing delays and improving system predictability, while optimizing the utilization of resources and reducing environmental impact. This effort is currently one of the five priority measures in the Flight Efficiency Plan published by IATA, CANSO and Eurocontrol \cite{eurocontrol}. In the U.S., the CDM-based ground delay program planning and control appeared in 1998; the stakeholders are the Joint Government Industry program, airlines, the Federal Aviation Administration including Air Traffic Control and Air Traffic Flow Management, and airports. The mechanisms involve the provision of accurate data (estimates of arrival and departure times) to stakeholders, the share of information, the airline decision to cancel or delay flights, and the rescheduling with priority constraints. Several improvements have been reported resulting from the CDM initiative, such as the Collaborative Departure Queue Management strategy at Memphis International Airport \cite{brinton2011collaborative} or the Surface Congestion Management scheme at New York's John F. Kennedy International Airport \cite{nakahara2011analysis}. However, there still is a growing need for more efficient and more robust tools to handle operations at airports. This effort should be combined with a necessary shift towards a better understanding of passengers' interests and concerns.

Most air travelers have experienced walking long distances in a passenger terminal to catch a flight or waiting on board their aircraft while it is waiting for a gate or is delayed by the movement of another aircraft. Many of such situations can be resolved or reduced by proper gate scheduling or assignment. 

\begin{figure}[tb]
	\centering
	\includegraphics[width=0.5\textwidth]{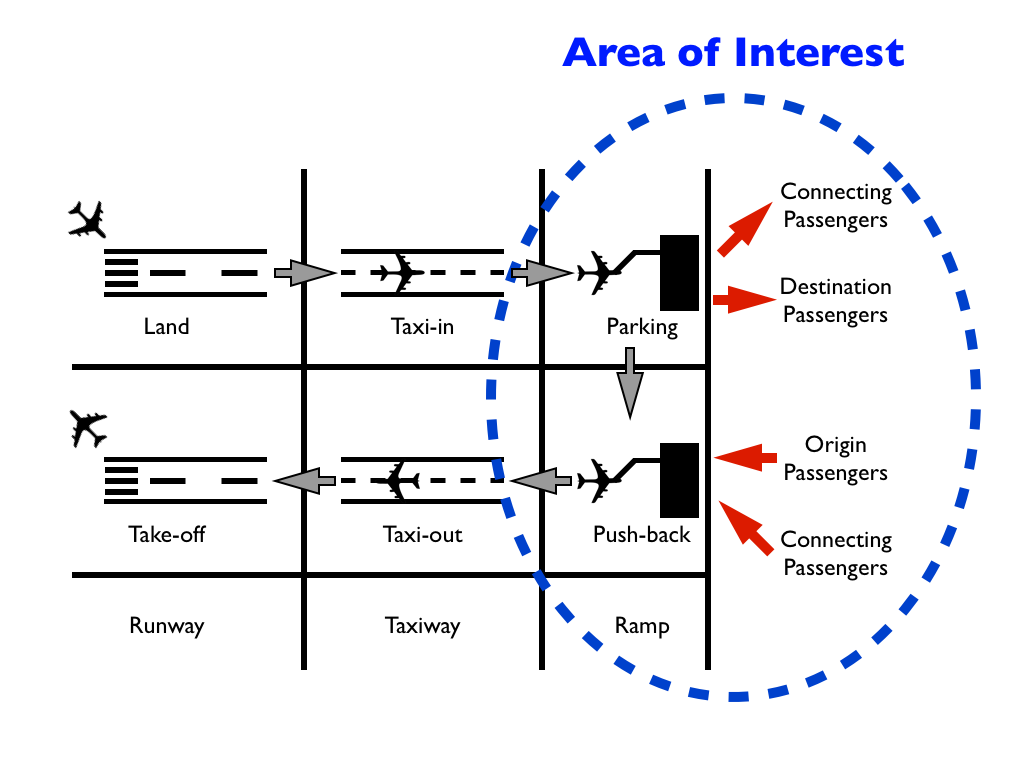}
	\caption{Airport operations and the focus of this study.}
	\label{f:airportOps}
\end{figure}

Airport operations range from landing to take-off of an aircraft as shown in Fig.~\ref{f:airportOps}. When an aircraft lands, it taxies into a ramp area and parks at a gate. While the aircraft is docking at the gate, passengers disembark and board the plane. When the aircraft is ready to depart, it pushes back and taxies out to a runway. Then, the aircraft takes off the airport. Among these operations, this study focuses on the optimization of ramp operations and the accommodation of passengers.

The first objective of this study is to minimize the transit time of passengers in a passenger terminal. The transit time of passengers consists of the time from the security checkpoint to a gate, from a gate to baggage claim, and from one gate to another gate. This is the most common objective of traditional studies focusing on gate assignment \cite{mangoubi1985oga, haghani1998oga}.

The second objective of this study is to minimize taxi time on ramps \cite{kim2012aga}. The taxi time depends on the length of taxi route. However, interfering taxi routes cause taxi delay. If two aircraft taxi in opposite directions on the same taxi lane, which does not happen on taxiway, it results in taxi delays. Because the taxi route of an aircraft is determined by the locations of assigned runway and gate, gate assignment is critical to reduce taxi time and taxi delays on ramps. 

The last objective of this study is to minimize disturbances in gate operations or equivalently to maximize the robustness of gate assignment \cite{bolat2000ppr, kim2011rga}. “Robust” means that the gate assignment is less sensitive to uncertain delays. Severe delays perturb gate operations by forcing arriving aircraft to wait for gates, or air traffic controllers to reassign gates. The disturbances can be reduced if the gate assignment is robust against uncertain delays. In addition, a robust gate assignment allows air traffic controllers to utilize gate-holding departure control more efficiently \cite{kim2012impact}. The gate-holding departure control delays push-backs in order to reduce taxi times and emission when the airport surface is congested. As a result, aircraft occupy gates longer than scheduled and it can negatively impact gate operations. If the gate assignment is robust, aircraft are able to stay longer at gates without disturbing gate operations and gate-holding departure control performs better.

All three objectives cannot be satisfied at the same time. Hence, this study presents trade-offs between objectives using flight schedules in a U.S. hub airport.

\section{Gate Assignment Problem}
\subsection{Data Source}
Previous studies on gate assignment generated fictitious passenger data (e.g., number of transfer passengers) because such data are not published. Thanks to a major U.S. carrier, this study is able to assign airport gates and analyze gate assignments with the actual number of transfer passengers at a U.S. major hub airport. Flight schedules and transfer passenger data on May 1st, 2011 at the hub airport are obtained from the carrier. All the flights are assumed to be full with passengers. Passengers who check in at the airport (origin passengers) and whose final destination is the airport (destination passengers) move from the passenger terminal to a gate. Passengers who have connecting flights at the airport (transfer passengers) move from a gate to another gate. Because only the data on the number of transfer passengers of a single carrier are available, passengers except transfer passengers among the carrier's flights are dealt with origin and destination (O\&D) passengers.

\subsection{Objective 1: Minimize Passenger Transit Time}
The first objective is to minimize the transit time of passengers. Passengers in an airport are categorized into three groups. Origin passengers begin their itinerary from the airport. Destination passengers finish their itinerary at the airport. Transfer passengers change from one flight to another at the airport.

The transit time of origin passengers depends on the distance from a security checkpoint to a gate ($d^{\mbox{s}}$). Let $v^{\mbox{m}}$ denote the average moving speed, which varies with the configuration of passenger terminal: $v^{\mbox{m}}$ is higher where passengers can move faster by taking moving sidewalk, underground people mover, etc. Assume that flight $i$ is assigned to gate $j$ and the number of origin passengers of flight $i$ is $n^{\mbox{o}}_i$, then the total transit time of origin passengers of flight $i$ is $n^{\mbox{o}}_i d^{\mbox{s}}_j/v^{\mbox{m}}$. Similarly, the total transit time of destination passengers of flight $i$ is $n^{\mbox{d}}_i d^{\mbox{b}}_j/v^{\mbox{m}}$, where $n^{\mbox{d}}_i$ is the number of destination passengers of flight $i$, and $d^{\mbox{b}}_j$ is the distance from gate $j$ to a baggage claim. Therefore, the transit time of O\&D passengers is determined by the location of a single gate because the locations of the security checkpoint and baggage claim are fixed.

Contrarily, the transit time of transfer passengers depends on the distance between two gates ($d_{jl}$). Let $n_{ik}$ denote the number of transfer passengers between flight $i$ and flight $k$. Then, the total transit time of passengers who transfer between flight $i$ and flight $k$ is $n_{ik} d_{jl}/v^m$.

Consequently, the transit times of O\&D passengers are expressed by linear terms and the transit times of transfer passengers are expressed by quadratic terms in the objective function (\ref{e:obj1}), where $x_{ij}$ is a decision variable that indicates whether flight $i$ is assigned to gate $j$. The formulation of the first objective is given below. 

\begin{align}
	\label{e:obj1}
	&Obj_{pax} = \mbox{minimize} \sum_{i \in \mathcal{F}} \sum_{j \in \mathcal{G}} (n^{\mbox{o}}_i \frac{d^{\mbox{s}}_j}{v^{\mbox{m}}} + n^{\mbox{d}}_i \frac{d^{\mbox{b}}_j}{v^{\mbox{m}}}) \ x_{ij} \nonumber\\
	&+ \sum_{i \in \mathcal{F}} \sum_{j \in \mathcal{G}} \sum_{k \in \mathcal{F}, k > i} \sum_{l \in \mathcal{G}} n_{ik} \frac{d_{jl}}{v^{\mbox{m}}} \ x_{ij} \ x_{kl} \\
	&\nonumber \text{subject to} \\
	\label{e:gateconst}
	&\sum_{j \in \mathcal{G}} x_{ij} = 1, \ \forall i \in \mathcal{F} \\
	\label{e:feasconst}
	&(t^{\mbox{out}}_i - t^{\mbox{in}}_k + t^{\mbox{buff}}) (t^{\mbox{out}}_k - t^{\mbox{in}}_i + t^{\mbox{buff}}) \nonumber\\
	&\leq M (2 - x_{ij} - x_{kj}), \ i \neq k, \ \forall i, k \in \mathcal{F}, \ \forall j \in \mathcal{G} \\
	\label{e:decision}
	&x_{ij} \in \{0,1\}, \ \forall i \in \mathcal{F}, \ \forall j \in \mathcal{G} \\
	&\nonumber \text{where } x_{ij} = \left\{ 
	\begin{array}{rl}
		1 & \text{if } f_{i} \text{ is assigned to }g_{j} \\
		0 & \text{otherwise.}
	\end{array} \right.
\end{align}

Two constraints are given in (\ref{e:gateconst})-(\ref{e:feasconst}). Equation (\ref{e:gateconst}) makes sure that every flight is assigned to a single gate. Equation (\ref{e:feasconst}) constrains two successive gate schedules, so that they are separated more than a certain amount of time, which is called buffer time ($t^{\mbox{buff}}$). Equation (\ref{e:feasconst}) is meaningful only if flights $i$ and $k$ are assigned to gate $j$ ($x_{ij}=x_{kj}=1$) because $M$ is an arbitrarily large number. $t^{\mbox{in}}_i$ indicates the scheduled gate-in time (arrival time) of flight $i$, $t^{\mbox{out}}_i$ indicates the scheduled gate-out time (departure time). $\mathcal{F}$ and $\mathcal{G}$ denote the sets of flights and gates. 

\subsection{Objective 2: Minimize Aircraft Taxi Time}
The second objective is to minimize unimpeded taxi time and taxi delay. The unimpeded taxi time for an arrival measures from when an aircraft enters a spot to when the aircraft parks at a gate without any taxi delay. The taxi time from a spot to a gate is calculated by dividing the distance from a spot to a gate by the taxi speed. The unimpeded taxi time for a departure measures from when an aircraft pushes back to when the aircraft leaves the ramp area without any taxi delay. It contains the duration to push back. A taxi delay happens in either of the following cases. 1) A taxiing aircraft blocks the push back route of another aircraft. 2) Two aircraft taxi in opposite directions on the same taxi lane. The first case is called a push back blocking and the push back is delayed until the taxiing aircraft passes through the push back route. The second case is called a taxi blocking and one of the aircraft must shift its taxi lane to another taxi lane. Therefore, the taxi routes of two aircraft condition taxi delays. 

Let $n^{\mbox{in}}_i$ denote the number of arrival passengers of flight $i$ and $u^{\mbox{in}}_j$ denote the unimpeded arrival taxi time of gate $j$. Note that arrival passengers include both destination passengers and transfer passengers of arriving flight $i$. Let $n^{\mbox{out}}_i$ denote the number of departure passengers of flight $i$ and $u^{\mbox{out}}_j$ denote the unimpeded departure taxi time of gate $j$. Then, the weighted unimpeded taxi time of flight $i$, which is weighted by the number of passengers on board, is $n^{\mbox{in}}_i u^{\mbox{in}}_j + n^{\mbox{out}}_i u^{\mbox{out}}_j$. Similar to the transit time of O\&D passengers of objective 1, the weighted unimpeded taxi time is expressed by a linear term in the objective function (\ref{e:obj2}).

Taxi delay ($t^{\mbox{dly}}$) involves a pair of aircraft and it is weighted by the sum of the number of passengers on board of both aircraft. For instance, if the taxi delay occurs between two arrivals, the total number of passengers is $n^{\mbox{in}}_i +n^{\mbox{in}}_k$. The quadratic terms of the objective function (\ref{e:obj2}) are weighted by a general form, $n_i +n_k$.

The formulation of the second objective is given below. The constraints of the first objective are applied equally.

\begin{align}
	\label{e:obj2}
	&Obj_{taxi} = \mbox{minimize} \sum_{i \in \mathcal{F}} \sum_{j \in \mathcal{G}} (n^{\mbox{in}}_i u^{\mbox{in}}_j + n^{\mbox{out}}_i u^{\mbox{out}}_j)  \ x_{ij} \nonumber\\
	&+ \sum_{i \in \mathcal{F}} \sum_{j \in \mathcal{G}} \sum_{k \in \mathcal{F}, k > i} \sum_{l \in \mathcal{G}} (n_{i} + n_{k}) t^{\mbox{dly}} \ x_{ij} \ x_{kl} \\
	&\nonumber \text{subject to (\ref{e:gateconst})-(\ref{e:decision}).}
\end{align}

\subsection{Objective 3: Maximize the Robustness of Gate Assignments}
\begin{figure}[tb]
	\centering
	\includegraphics[width=0.5\textwidth]{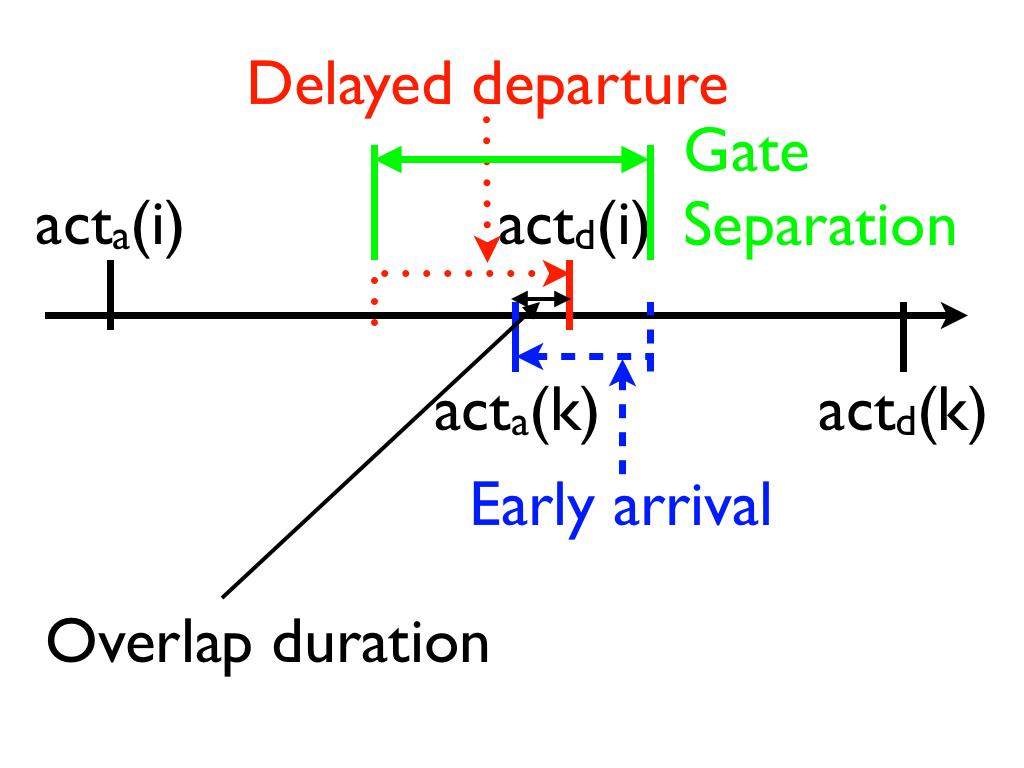}
	\caption{Gate conflict and its duration.}
	\label{f:gateconflict}
\end{figure}

The third objective is to maximize the robustness of gate assignments. Equivalently, the objective is to minimize the duration of gate conflicts. If a gate is still occupied by an aircraft when another aircraft requests the gate, the latter should wait until the assigned gate or another gate is available, which corresponds to a gate conflict. Fig.~\ref{f:gateconflict} illustrates a gate conflict, where $act_a(i)$ and $act_d(i)$ denote the actual arrival and departure times of flight $i$, and the gate separation is the time gap between the scheduled departure time of flight $i$ ($t^{\mbox{out}}_i$) and the scheduled arrival time of flight $k$ ($t^{\mbox{in}}_k$). In Fig.~\ref{f:gateconflict}, flight $i$ is scheduled to leave the gate before flight $k$ arrives, but the departure time of flight $i$ is delayed and flight $k$ arrives earlier than schedule. So, when flight $k$ arrives, the gate is not released yet and flight $k$ has to wait for a gate.

Because the actual arrival and departure times are unknown when gates are assigned, the duration of a gate conflict is estimated based on the probability distributions of arrival delay and departure delay. The expected duration of a gate conflict is calculated by $E[act_d(i)-act_a(k) | act_d(i)>act_a(k)]$ when $t^{\mbox{in}}_k > t^{\mbox{out}}_i$. Details of the calculation are given in \cite{kim2011rga}.

\begin{figure}[tb]
	\centering
	\includegraphics[width=0.5\textwidth]{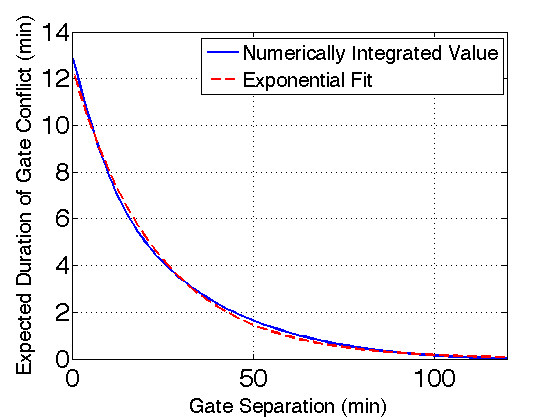}
	\caption{Expected duration of gate conflict.}
	\label{f:duration}
\end{figure}

The expected duration of a gate conflict is known to rely on gate separation \cite{kim2011rga}. Using the delay data of a U.S. carrier at a hub airport in May 2011, the expected duration is shown in Fig.~\ref{f:duration} and it is fit to $a\times b^{sep(i,k)}$, where $a$ and $b$ are constants and $sep(i,k)$ denotes the gate separation between flights $i$ and $k$.

The formulation of the third objective is given below. Note that the expected duration of a gate conflict is weighted by the number of arrival passengers. 

\begin{align}
	\label{e:obj3}
	&Obj_{robust} = \mbox{minimize} \sum_{i \in \mathcal{F}} \sum_{k \in \mathcal{F}, k > i} n^{\mbox{in}} a\times b^{sep(i,k)} \sum_{j \in \mathcal{G}} \ x_{ij} \ x_{kj} \\
	&\nonumber \text{subject to (\ref{e:gateconst})-(\ref{e:decision}).}
\end{align}

\subsection{Trade-offs of Multiple Objectives}
It is known that there are trade-offs between objectives that are presented in this study \cite{kim2012aga, kim2013rga}. In order to analyze the trade-offs between three objectives, five scenarios are presented in Table~\ref{t:scenarios}. Scenarios 1-3 optimize each objective. Scenario 4 weighs equally on objective 1 and 2. So, it optimizes two objectives concurrently. Scenario 5 takes all the objectives into account: 40\% on the first and second objectives and 20\% on the third objective. These percentages are chosen arbitrarily: the proportion of three objectives depends on the policy of airport gate managers and airlines.

\begin{table*}[tb]
	\begin{center}
	\caption{Scenarios}
	\label{t:scenarios}
	\begin{tabular}{l|l|l}
		Scenario & Objective Function & Explanation \\ \hline
		Scenario 1 & $Obj = Obj_{pax}$ & Optimize objective 1 only.\\
		Scenario 2 & $Obj = Obj_{taxi}$ & Optimize objective 2 only.\\
		Scenario 3 & $Obj = Obj_{robust}$ & Optimize objective 3 only.\\
		Scenario 4 & $Obj = 0.5 Obj_{pax} + 0.5 Obj_{taxi}$ & Balance objective 1 and 2.\\
		Scenario 5 & $Obj = 0.4 Obj_{pax} + 0.4 Obj_{taxi} + 0.2 Obj_{robust}$ & Balance all the objectives.
	\end{tabular}
	\end{center}
\end{table*}

\subsection{Optimization Method}
\begin{figure}[tb]
	\centering
	\includegraphics[width=0.5\textwidth]{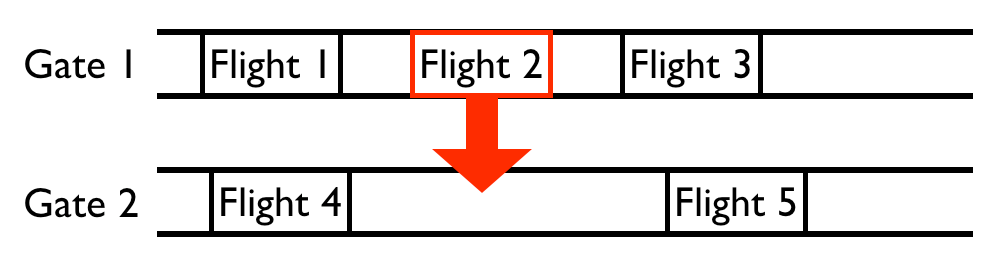}
	\caption{Insert move.}
	\label{f:insert}
\end{figure}

\begin{figure}[tb]
	\centering
	\includegraphics[width=0.5\textwidth]{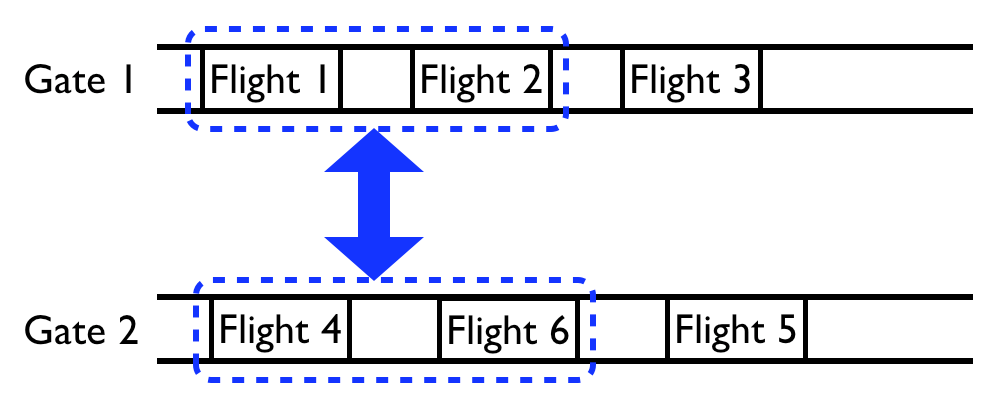}
	\caption{Exchange move.}
	\label{f:exchange}
\end{figure}

The Tabu Search (TS) is a meta-heuristic algorithm known to efficiently deal with combinatorial optimization problems such as the gate assignment problem \cite{glover1997tabu, xu2001aga}. Because the gate assignment problem is complex, it is hard to find the optimal solution in a reasonable time. The TS can outperform the Branch and Bound and Genetic Algorithm in terms of solution time and solution accuracy for the gate assignment problem \cite{kim2012aga}. The TS is a local search so the algorithm can converge to a local optimum, which is not the global optimum. In order to help the TS escape from a local optimum, a tabu memory restricts the TS from utilizing recently used search moves for certain iterations. However, if a restricted search move improves the objective value, the search move can be used regardless of the tabu memory, known as the aspiration criterion. Two types of neighborhood search moves of TS are shown in Fig.~\ref{f:insert} and Fig.~\ref{f:exchange}. The insert move changes a flight's gate assignment from one to another, and the interval exchange move swaps the gate assignments of two groups of flights. 

The TS iterates until the number of iteration reaches the maximum iteration or there is no improvement of the objective value after some iterations past the last best score. The insert move is evaluated at every iteration in order to intensify a local search around a narrow neighborhood of the current solution. The interval exchange move is evaluated periodically in order to diversify the search: the interval exchange move brings a relatively large change in the current solution. More details of the implementation of the TS on the gate assignment problem are given in \cite{kim2012aga}.

\section{Results}
\begin{figure}[tb]
	\centering
	\includegraphics[width=0.5\textwidth]{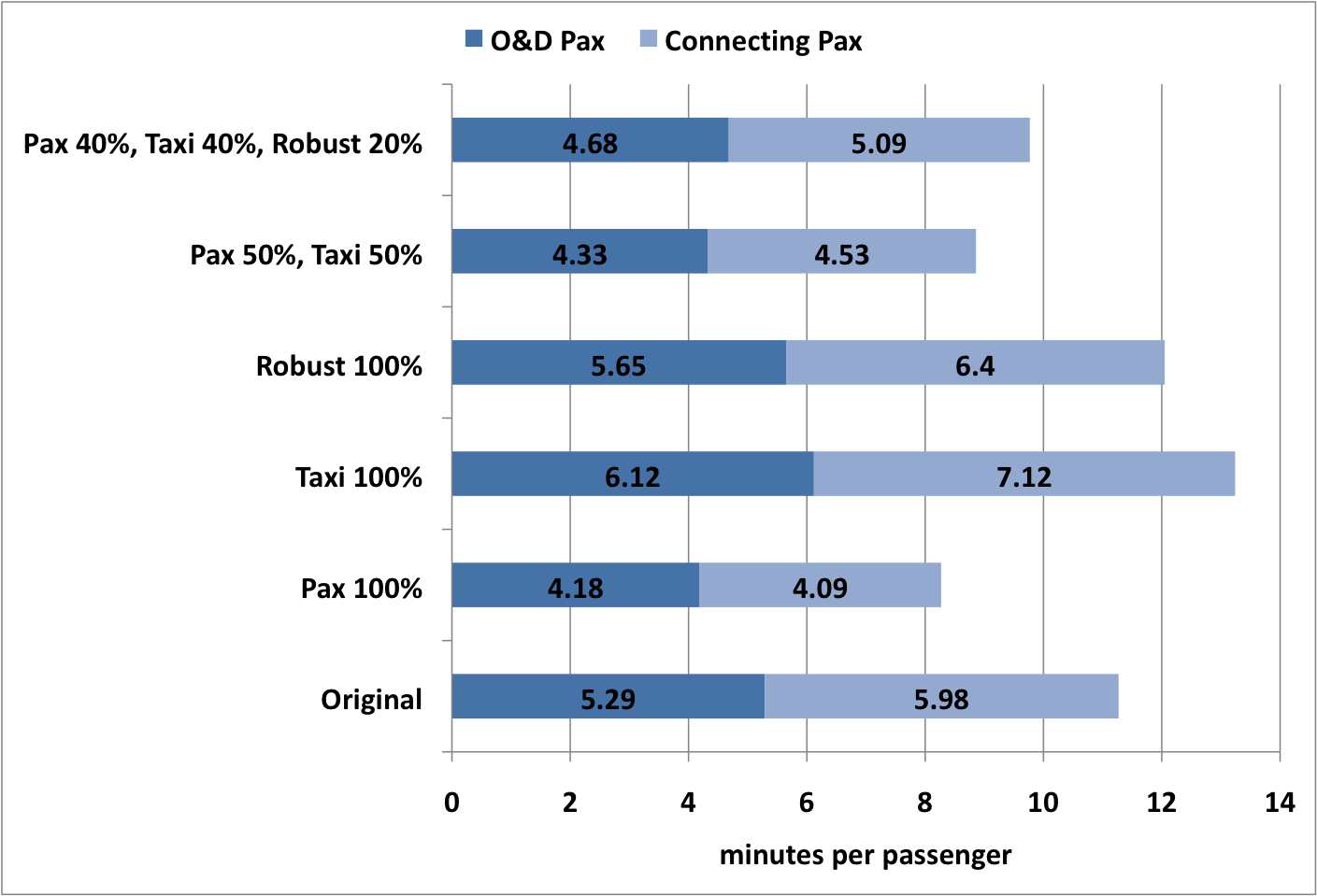}
	\caption{Transit time per passenger in minute.}
	\label{f:transitresult}
\end{figure}

Fig.~\ref{f:transitresult} illustrates the average transit time of each passenger. As expected, scenario 1 (Pax 100\%) results in the shortest passenger transit time. In scenarios 4 and 5, passengers walk longer than in scenario 1 but less than in the original (current) gate assignment and scenarios 2 and 3.

\begin{figure}[tb]
	\centering
	\includegraphics[width=0.5\textwidth]{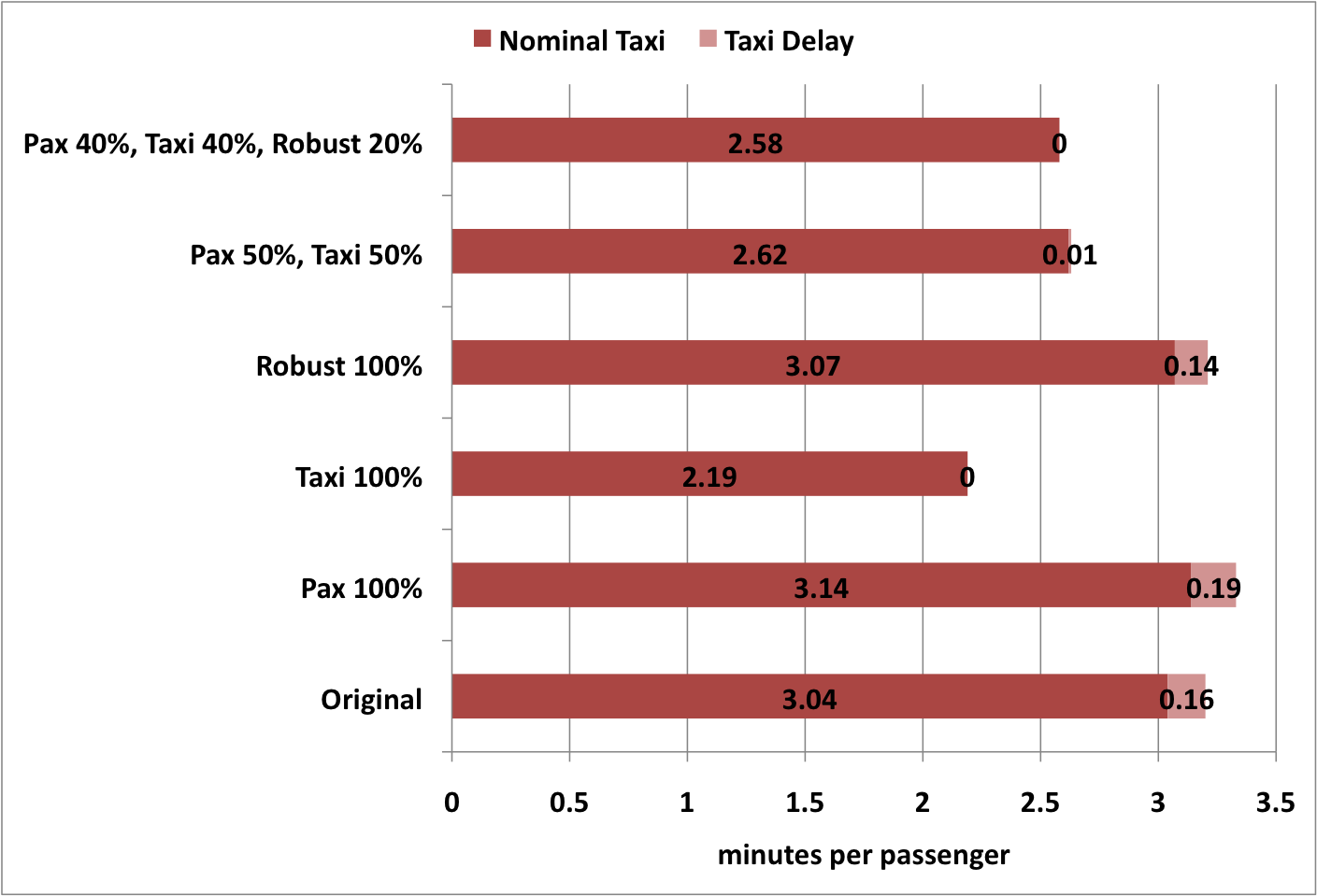}
	\caption{Aircraft taxi time per passenger in minute.}
	\label{f:taxiresult}
\end{figure}

Fig.~\ref{f:taxiresult} shows the average taxi time of each passenger. Undoubtedly, scenario 2 (Taxi 100\%) produces the shortest taxi time with zero taxi delay. From Fig.~\ref{f:transitresult} and Fig.~\ref{f:taxiresult}, it is inferred that scenarios 4 and 5 balance well between two different objectives: minimizing passenger transit time and minimizing taxi time.

\begin{figure}[tb]
	\centering
	\includegraphics[width=0.5\textwidth]{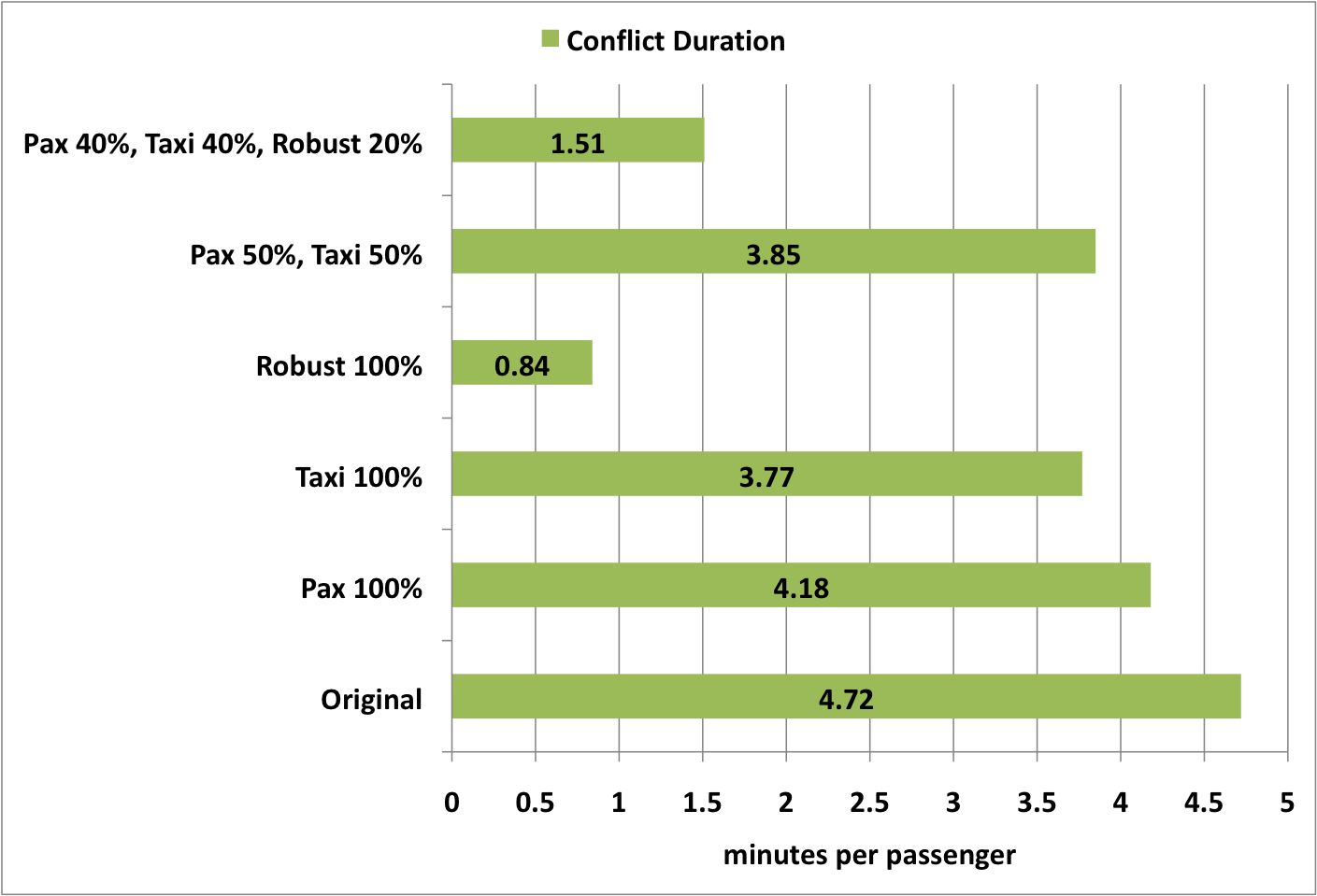}
	\caption{Expected duration of gate conflict per passenger in minute.}
	\label{f:robustresult}
\end{figure}

Fig.~\ref{f:robustresult} shows the average gate conflict duration of each passenger. Scenario 3 (Robust 100\%) induces the shortest duration of gate conflict. Note that scenario 5 (Pax 40\%, Taxi 40\%, Robust 20\%) gives the second best result of the robustness of gate assignment while scenario 4 (Pax 50\%, Taxi 50\%) gives less robust gate assignment. Consequently, scenario 5 balances three objectives at the same time better than other scenarios.

From the results, scenarios 1, 2, and 3 give the best results for each objective but lead to poor performance with respect to other objectives. Also, scenario 4 performs well just for objective 1 and 2. Only scenario 5 succeeds at satisfying all the objectives. The original gate assignment, however, does not show satisfactory performance for the given objectives. 

\section{Conclusion}
This study presented several simulations of gate assignments according to different objectives. These objectives are minimizing transit time of passengers in passenger terminals; minimizing aircraft taxi time on ramps; and minimizing the duration of gate conflicts. It is known that there are trade-offs between these objectives so an objective function that balances three objectives at the same time was proposed. The balancing objective function satisfies all the three objectives while other objective functions only satisfy one or two objectives. Moreover, the balancing objective function outperforms the current gate assignment in every objective. Therefore, the gate assignment of the airport has a potential to be improved regarding the efficiency of traffic flow in passenger terminals and on ramps, as well as on the robustness of gate operations.

Future work will account for gate-holding strategies generated by Airport CDM. Although this study includes the robustness of gate assignment, which was shown to help gate-holding strategies perform better, a comprehensive analysis of gate-holding strategies and passengers' experience at the airport is still needed. Hence, future work will address the impact of gate-holding strategies on passengers.

\bibliography{atl}
\bibliographystyle{unsrt}

\section*{Acknowledgment}
This work was supported in part by the European Community Framework Programme 7 under the META-CDM project.

\section*{Author Biography}
\textbf{Sang Hyun Kim} is a Ph.D. candidate in the School of Aerospace Engineering at the Georgia Institute of Technology. He holds his B.S. degree in mechanical and aerospace engineering from Seoul National University, South Korea. His research interests are optimization, transportation, airport operations, ramp management, and general aerospace engineering. 

\textbf{Aude Marzuoli} is currently pursuing a PhD in Aerospace Engineering at the Georgia Institute of Technology, with a focus on air traffic management, optimization and control. In 2012, she obtained her Master in Aerospace Engineering from Georgia Tech and her Engineering Diploma from Supelec in France. She previously attended the Lycee Henri IV in Paris for her preparatory classes.

\textbf{Eric Feron} is received the B.S. degree from Ecole Polytechnique, Palaiseau, France, the M.S. degree from the Ecole Normale Supérieure, Paris, France, and the Ph.D. degree from Stanford University, Stanford, CA. He is the Dutton–Ducoffe Professor of Aerospace Software Engineering, Georgia Institute of Technology, Atlanta, GA. Prior to that, he was with the faculty of the Department of Aeronautics and Astronautics, Massachusetts Institute of Technology, Cambridge, for 12 years. His former research students are distributed throughout academia, government, and industry. He has published two books and several research papers. His research interests include using fundamental concepts of control systems, optimization, and computer science to address important problems in aerospace engineering such as aerobatic control of unmanned aerial vehicles and multiagent operations such as air traffic control systems and aerospace software system certification.

\textbf{John-Paul Clarke} is an Associate Professor in the Daniel Guggenheim School of Aerospace Engineering with a courtesy appointment in the H. Milton Stewart School of Industrial and Systems Engineering, and Director of the Air Transportation Laboratory at the Georgia Institute of Technology. He received S.B. (1991), S.M. (1992), and Sc.D. (1997) degrees in aeronautics and astronautics from the Massachusetts Institute of Technology. His research and teaching in the areas of control, optimization, and system analysis, architecture, and design are motivated by his desire to simultaneously maximize the efficiency and minimize the societal costs (especially on the environment) of the global air transportation system.

\textbf{Daniel Delahaye} is doing research for the French Air Navigation Research Center since 1995 and is member of the artificial evolution team of the applied mathematics research center (CMAP from Ecole Polytechnique, France). Delahaye graduated with an Engineer degree from ENAC, a Master of Science in signal processing from the national polytechnic institute of Toulouse in 1991, the PhD D in automatic control from Ecole Nationale Sup\'erieure de l'Aéronautique et de l'Espace (SUPAERO) in 1995. Following a Post-Doc in the Department of Aeronautics and Astronautics at MIT in 1996, he has been working continuously in the Applied Mathematic Laboratory of ENAC, where he is conducting research on stochastic optimization for large scale air traffic management.

\end{document}